# Random quasi-phase-matched second-harmonic generation in periodically poled lithium tantalate

S. Stivala,[1] A. C. Busacca,[1] A. Pasquazi,[2,*] R. L. Oliveri,[3] R. Morandotti,[2] and G. Assanto[4]

[1]*DIEET, University of Palermo, 90128 Palermo, Italy*
[2]*Ultrafast Optical Processing Group, INRS-EMT Université du Quebec, Varennes, Quebec J3X 1S2, Canada*
[3]*DIFTER, University of Palermo, 90128 Palermo, Italy*
[4]*NooEL—Nonlinear Optics and OptoElectronics Laboratory, CNISM, University "Roma Tre," 00146 Rome, Italy*
*\*Corresponding author: alessia.pasquazi@gmail.com*

We observe second harmonic generation via random quasi-phase-matching in a 2.0 $\mu$m periodically poled, 1-cm-long, $z$-cut lithium tantalate. Away from resonance, the harmonic output profiles exhibit a characteristic pattern stemming from a stochastic domain distribution and a quadratic growth with the fundamental excitation, as well as a broadband spectral response. The results are in good agreement with a simple model and numerical simulations in the undepleted regime, assuming an anisotropic spread of the random nonlinear component.

Parametric wavelength conversion in media with a random nonlinearity, discussed in the pioneering works by Freund [1] and by Dolino and coworkers [2], has recently attracted significant interest due to its nonresonant behavior, i.e., its low sensitivity to phase matching and its broadband response [3–9]; potential applications have also been investigated, such as the time-domain characterization of femtosecond pulses [10]. Moreover, in connection with quasi-periodicity, a few important aspects of nonlinear wave mixing and transverse phase matching have been unveiled [11–14]. Broadband frequency conversion can take place in both isotropic polycrystalline ferroelectrics [3] and in crystals with randomly distributed antiparallel ferroelectric micro-domains [5,6]; the former are easily accessible and support a linear trend in terms of the generated second harmonic (SH) versus the input fundamental power, while the latter exhibit the standard quadratic dependence of the generated SH versus the input fundamental power, with higher conversion efficiencies.

The presence of antiparallel ferroelectric domains can intrinsically depend on the specific growth technique, as is the case for strontium barium niobate (SBN, $d_{33}$=12 pm/V) [4]. Here, a random domain distribution is generated during boule preparation [7,11]. In SBN and with reference to SH generation, Horowitz and coworkers studied the role of microdomains superimposed to an ordered pattern [6]. In several ferroelectric crystals with a second-order susceptibility, however, electric field poling for quasi-phase matching (QPM) via domain inversion is the most used approach for engineering nonlinear gratings and realizing a variety of bulk and guidedwave structures for signal processing based on frequency conversion [15,16]. An intense electric field applied across the thickness of a ferroelectric crystal relates to the probability of domain nucleation. Domain nucleation, however, is a stochastic process with a random trend due to various defects, including vacancies, impurities, ions in wrong lattice sites, etc. [17]. The random character becomes particularly relevant when the imposed (periodic) pattern has small features, with the final structure being prone to show a disordered two-dimensional distribution of mark-to-space-ratio (MTSR) superimposed to the grating defined by the electrode geometry.

In this Letter we report on second-harmonic generation (SHG) via random quasi-phase-matching (rQPM). Our experiments were carried out in a bulk periodically poled lithium tantalate (PPLT) with a random MTSR over a propagation length of 1 cm. While the SHG at phase matching with the QPM grating was not affected by the presence of random domains, far from the QPM-SHG resonant wavelength the harmonic signal was remarkably larger than that generated in the unpoled portions of the sample, with an output SH profile strongly dependent on both the wavelength and the position of the input beam.

We prepared the rQPM samples from optical grade wafers of a $z$-cut 500-$\mu$m-thick congruent lithium tantalate (LT, $d_{33}$=20 pm/V) [18]. To implement the periodic poling, we applied high-voltage pulses across the LT thickness using an electrolyte gel and an insulating mask. After spin coating the $-z$ sample facet with a 1.3-$\mu$m-thick photoresist film, we defined a 2.0 $\mu$m period grating with grooves parallel to the $y$ axis of the crystal by way of standard photolithography. In order to exceed the LT coercive field we applied a single 1.3 kV pulse superimposed to a 10 kV bias for a time interval (300 $\mu$s) long enough to obtain a periodic inversion over the entire patterned area with a size of 0.7 mm$\times$10.0 mm. After poling, some of the samples were etched in hydrofluoric acid for several minutes to reveal the domain patterns at the $-z$ surface as shown in Fig. 1(a): imaging via scanning electron microscopy clearly shows the random MTSR in the QPM grating. The poling dynamics is such that the first enucleated (inverted) domains spread in the $xy$ crystal plane and become wider than those generated later. Despite the dominant role of/

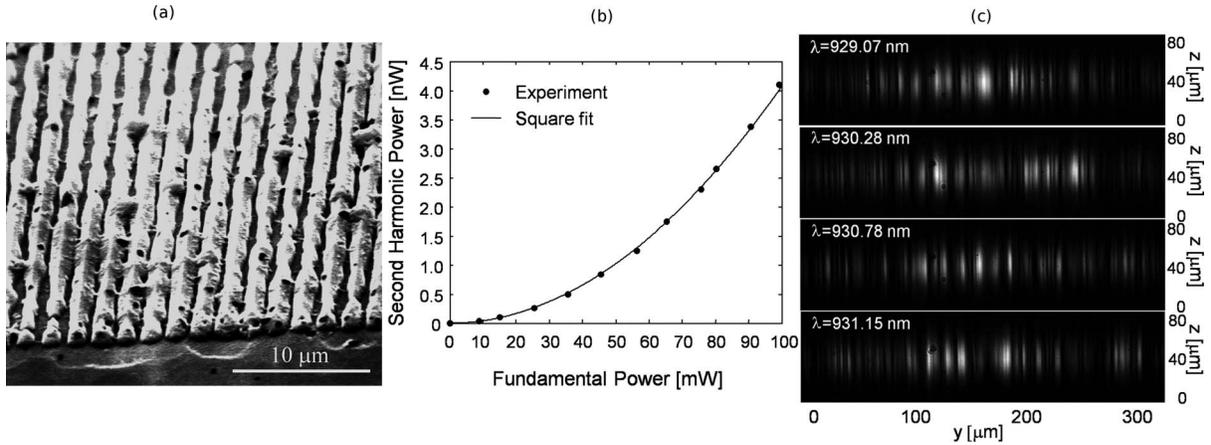

Fig. 1. (a) Scanning electron micrograph of an etched LT sample after periodic poling. (b) Measured SH generated versus FF input power at $\lambda_{FF}$=929.66 nm. (c) Acquired near-field SH intensity patterns due to rQPM: the FF input wavelengths are indicated in the legends.

the 2.0 μm periodic electrode pattern in defining the ferroelectric grating (as demonstrated by the first-order QPM-SHG on similar substrates [19]), the final MTSR has nonuniform stochastic features superimposed to the regular pattern.

For the nonlinear characterization, the fundamental frequency (FF) $TEM_{00}$ input beam was provided by a cw Ti:sapphire laser, tunable from λ = 700 to 980 nm with 40 GHz linewidth, and was focused to a waist of 25 μm into the 1-cm-long sample, held at 195 °C to reduce the risk of photorefractive damage. The generated SH was spectrally filtered at the output and measured with a photomultiplier and a lock-in amplifier. Both the FF and SH near- and far-field intensity output profiles were acquired via a high resolution CCD camera with and without a 10× microscope objective, respectively.

When exciting the sample with a 724 nm FF, we observed a bell-shaped SH across the whole inverted substrate. This is a consequence of the first-order QPM in bulk, due to the 2.0 μm periodicity in the $d_{33}$ quadratic nonlinearity; however, detunings as small as 1 nm, i.e., from $\lambda_{FF} \approx$ 724 to 725 nm, abruptly caused the SH level to drop down to the noise background level. The rQPM contribution manifested at FF wavelengths above 900 nm, where we could observe a larger SH power increasing with $\lambda_{FF}$: its parabolic trend versus the input FF power at $\lambda_{FF}$ = 929.66 nm is visible in Fig. 1(b). This rQPM-SHG conversion efficiency is about 1 order of magnitude lower than that for a fundamental wavelength of 724 nm. The near-field CCD images of the SH showed a random set of finger-like spots, parallel to one another and to the z axis, with intensities and distributions strongly depending on the location and the wavelength of the input beam. Figure 1(c) displays a selection of typical photographs, taken by varying slightly the input FF wavelength. The central wavelength value was monitored by way of an optical spectrum analyzer. The use of a cw source allowed us to record much sharper SH patterns than those otherwise obtainable with a typical pulsed-laser [5]. On the other hand, the SH generated outside the poled region (i.e., in a homogeneous bulk LT) maintained the FF Gaussian shape, with conversion efficiencies more than 2 orders of magnitude lower than in the rQPM case and with a response that was substantially wavelength independent in the available laser spectral range.

The nonresonant trend of the SHG is characteristic of the rQPM. While in the standard QPM-SHG the interaction occurs via one reciprocal lattice vector, in nonlinear disordered media, such as the PPLT with a nonuniform MTSR, the phase mismatch can be compensated by several such vectors [2,7], supporting a broadband generation. The fringes visible in the SHG pattern are parallel to z, indicating negligible variations along z of the domain dishomogeneity. This effect can be numerically simulated with a beam propagator, solving the coupled equations for degenerate upconversion and assuming a stochastic distribution for the nonlinearity. With reference to the PPLT and to the rectangular modulation ($\pm d_{33}$) of the quadratic coefficient $d_{33}$, we superimposed a white noise in the plane xy to a one-dimensional grating with $\Lambda$ = 2.0 μm along x. Such a noise component was suitably filtered in the Fourier domain ($k_x, k_y$), taking the superposition of two Gaussians along $k_x$ and a decaying exponential along $k_y$ [see Fig. 2(a)], yielding the spatial pattern shown in Fig. 2(b). This heuristic model for the random MTSR on a short-period QPM matched remarkably well the experimental results: at the QPM-SHG resonance [Fig. 2(c)], the generated harmonic is substantially unaffected by the randomly distributed nonlinearity. Conversely, far from the QPM [Fig. 2(d)], the frequency-doubled signal spreads in propagation with the characteristic fingerprint of the SHG via the rQPM [4].

Figures 3(a) and 3(b) show the measured and simulated far-field SH distributions along y, whereas Figs. 3(c) and 3(d) display the actual and calculated conversion efficiencies versus the FF wavelength in the whole 900–990 nm interval explored in the characterization. The filter profile [Fig. 2(a)], adjusted to fit the experimental data, is the characteristic "signature" of the random process and shows that the noise in the "domain size" is widely distributed in the xy plane, with larger features than the QPM period;

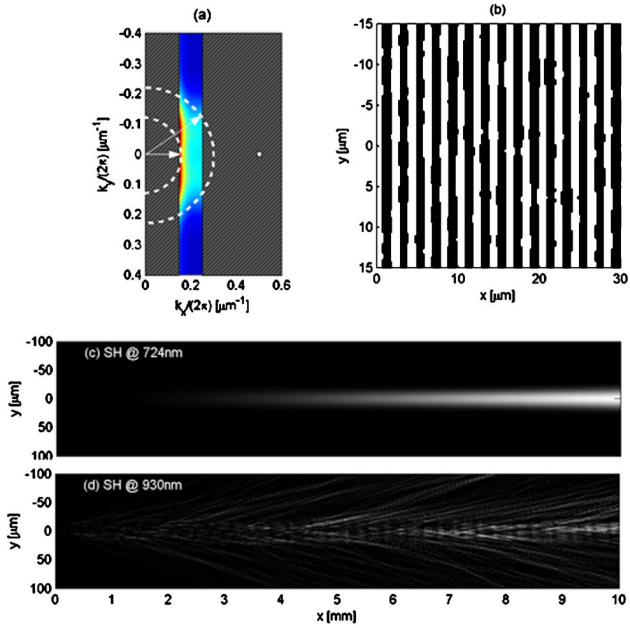

Fig. 2. (Color online) (a) Pseudocolor map of the Fourier filter, with two Gaussians along $k_x$ and a decaying exponential along $k_y$. The white dot indicates the $\Lambda=2.0$ $\mu$m QPM component. The uniform (gray dashed) region is outside the FF range of wavelengths used in this investigation. The arrows point to the longest and shortest Fourier components, corresponding to average domain features in the interval 3.5–6.5 $\mu$m. (b) Random distribution of the sign of the quadratic coefficient $d_{33}$ in the $xy$ plane; (c) SH evolution in $xy$ at the QPM-SHG resonance (724 nm); (d) SH evolution in $xy$ for a FF wavelength (930 nm) well away from the QPM resonance.

from the profile we can infer random domains with a distribution in the length range 3.5–6.5 $\mu$m, although the increase in the conversion toward longer wavelengths [Fig. 3(c)] pinpoints the presence of longer domain features. The noise distribution is substantially uncorrelated with the mask for periodic poling, consistent with the random nucleation of inverted domains.

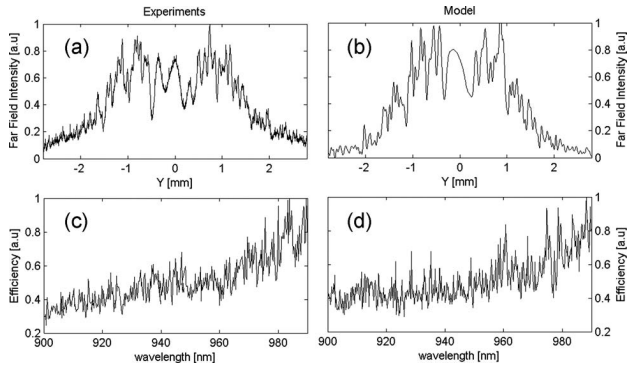

Fig. 3. Comparison between experimental data and numerical simulations. Top panels, far-field intensity distribution (a) acquired by the CCD at $\lambda_{FF}=929.89$ nm and averaged over 50 input locations, and (b) calculated at the same $\lambda_{FF}$ and averaged over 20 noise realizations. Bottom panels, (c) measured and (d) calculated SHG conversion efficiencies versus FF wavelength.

In conclusion, we have observed and characterized the SH generation in random quasi-phase-matched short-period LT structures prepared by electric field poling. The frequency-doubling process resulted to be significantly broadband, with a rQPM generated SH power more than 2 orders of magnitude higher than in the bulk unpoled LT. The output SH profile, as well as its dependence on the FF wavelength and power, was described with the aid of a stochastic distribution of the nonlinear domains, yielding a random component with a domain size between 3.5 and 6.5 $\mu$m.

The work was funded by the Italian Ministry for Scientific Research through Project PRIN 2007CT355 coordinated by S. Riva Sanseverino and the Natural Sciences and Engineering Research Council of Canada (NSERC). A. Pasquazi is supported by "Le Fonds québécois de la recherché sur la nature et les technologies" through the Ministère de l'Éducation, du Loisir et du Sport fellowship.